\documentclass[aps,twocolumn,prd,showpacs,amsmath,amssymb]{revtex4}
\usepackage{dcolumn}
\usepackage{bm}
\usepackage{amsfonts}

\newcommand{\be}{\begin{equation}}
\newcommand{\ee}{\end{equation}} 
\newcommand{\eei}{\end{equation}\indent\indent}
\newcommand{\bc}{\begin{center}}
\newcommand{\ec}{\end{center}}
\newcommand{\ber}{\begin{eqnarray*}}
\newcommand{\ear}{\end{eqnarray*}}
\newcommand{\ba}{\begin{array}}
\newcommand{\ea}{\end{array}}

\newcommand{\bea}{\begin{eqnarray}}
\newcommand{\eea}{\end{eqnarray}}

\newcommand{\ei}{\end{itemize}}

\newcommand{\bra}[1]{\left(#1\right)}
\newcommand{\bras}[1]{\left[#1\right]}
\newcommand{\brac}[1]{\left\{#1\right\}}

\newcommand{\nab}{\nabla}
\newcommand{\la}{\langle}
\newcommand{\ra}{\rangle}

\newcommand \veps {\varepsilon} 

\newcommand{\lb}{\{}
\newcommand{\rb}{\}}
\newcommand{\A}{{\cal A}}
\newcommand{\E}{{\cal E}}
\renewcommand{\H}{{\cal H}}

\newcommand{\lc}{\varepsilon}

\def\case#1/#2{\textstyle\frac{#1}{#2} }

\newcommand{\hatn}{a}
\newcommand{\dotn}{\alpha}
\newcommand{\udota}{{\cal A}}
\newcommand{\f}[2]{\textstyle\frac{#1}{#2}}

\begin{document}

\title{Almost Birkhoff Theorem in General Relativity}

\author{Rituparno Goswami and George F R Ellis}
\affiliation{ACGC and Department of Mathematics and Applied
Mathematics, University of Cape Town, Rondebosch, 7701, South
Africa} 

\date{\today}

\email{Rituparno.Goswami@uct.ac.za} \email{George.Ellis@uct.ac.za}

\begin{abstract}
We extend Birkhoff's theorem for almost LRS-II vacuum spacetimes to
show that the rigidity of spherical vacuum solutions of Einstein's
field equations continues even in the perturbed scenario.
\end{abstract}

\maketitle

\section{Introduction}
The core content of Birkhoff's theorem \cite{birk} is that any
spherically symmetric solution of the vacuum field equations has
an extra symmetry: it must be either locally static or spatially
homogeneous. This underlies the crucial importance in astrophysics
of the Schwarzschild solution, as it means that the exterior
metric of any exactly spherical star must be given by the
Schwarzschild metric (it cannot be Minkowski spacetime if the mass
is non-zero); and this also underlies the uniqueness results for
non-rotating black holes.

However it is an exact theorem that is only valid for exact
spherical symmetry; but no real star is exactly spherically
symmetric. So a key question is whether the result is approximately
true for approximately spherically symmetric vacuum solutions. We
prove an ``almost Birkhoff theorem'' in this paper that shows this
is indeed the case, so those results carry over to astrophysically
realistic situations (such as the Solar System). There are of course
many papers discussing perturbations of the Schwarzschild solution,
but none appear to focus on this specific issue. It is in a sense an
analogue of another important result, the ``almost EGS theorem''
\cite{almost EGS} that generalizes from exact isotropy of radiation
to approximate isotropy the crucial Ehlers-Geren-Sachs theorem
\cite{EGS}, proving that isotropic cosmic background radiation
everywhere in an expanding universe domain $U$ implies a
Robertson-Walker geometry in that domain. In both cases an exact
mathematical result, depending on exact symmetry, is generalized to
a more physically realistic result, depending on approximate
symmetry.

The almost-Birkhoff result will not be true if spacetime is not a
vacuum (empty) spacetime, for the degrees of freedom available
through a matter source generically invalidate the result, as is
shown for example by the family of Lema\^{\i}tre-Tolman-Bondi
(LTB) models \cite{LTB}. However it remains true for the
electrovac case \cite{electrovac} and solutions with a
cosmological constant. It will also not be true for gravitational
theories with a scalar degree of freedom, such as scalar-tensor
theories \cite{scalar-tensor}. The rigidity embodied in this
property of the Einstein Field Equations is specific to vacuum
General Relativity solutions, or solutions with a trace-free
matter tensor. One should also note that (like the EGS result) the
result is local: both Birkhoff's theorem, and our generalization
of it, are independent of boundary conditions at infinity: they
hold in local neighborhoods $U$. There are of course crucial
differences between the exactly spherical case and that of
approximate symmetry: specifically, an almost spherically
symmetric pulsating star can emit gravitational waves, but this is
not possible in the exactly spherical case. However we show that
in such a case if the intensity of the radiation is such as to
leave the solution almost spherically symmetric, the space time
will remain almost static, which is defined in a way we will make
precise later.

We prove the result by using the 1+1+2 covariant perturbation
formalism \cite{extension,chris}, which developed from the 1+3
covariant perturbation formalism \cite{Covariant}. This enable us
to prove the approximate result as a straightforward
generalization of the exact result, which we prove first, using
the 1+1+2 formalism. Actually we prove a small generalization of
the standard Birkhoff result: it holds for all Class II Locally
Rotational Symmetric (LRS) spacetimes \cite{EllisLRS} (which
include Schwarzschild as a special case).

\section{1+1+2 Covariant splitting of spacetime}
The 1+3 covariant formalism 
{\cite{Covariant}} has proven to be a very useful technique in many
aspects of relativistic cosmology. In this approach first we define
a timelike congruence by a timelike unit vector $u^a$ ($u^a u_a =
-1$). Then the spacetime is split in the form $R\otimes V$ where $R$
denotes the timeline along $u^a$ and $V$ is the tangent 3-space
perpendicular to $u^a$. Then  any vector $X^a$ can be projected on
the 3-space by the projection tensor $h^a_b=g^a_b+u^au_b$. The
vector $ u^{a} $ is used to define the \textit{covariant time
derivative} (denoted by a dot) for any tensor $ T^{a..b}{}_{c..d} $
along the observers' worldlines defined by \be
\dot{T}^{a..b}{}_{c..d}{} = u^{e} \nab_{e} {T}^{a..b}{}_{c..d}~, \ee
and the tensor $ h_{ab} $ is used to define the fully orthogonally
\textit{projected covariant derivative} $D$ for any tensor $
T^{a..b}{}_{c..d} $ , \be D_{e}T^{a..b}{}_{c..d}{} = h^a{}_f
h^p{}_c...h^b{}_g h^q{}_d h^r{}_e \nab_{r} {T}^{f..g}{}_{p..q}~, \ee
with total projection on all the free indices.

In the (1+1+2) approach we further split the 3-space $V$, by
introducing the spacelike unit vector $ e^{a} $ orthogonal to $
u^{a} $ so that \be e_{a} u^{a} = 0\;,\; \quad e_{a} e^{a} = 1. \ee
Then the \textit{projection tensor} \be N_{a}{}^{b} \equiv
h_{a}{}^{b} - e_{a}e^{b} = g_{a}{}^{b} + u_{a}u^{b} -
e_{a}e^{b}~,~~N^{a}{}_{a} = 2~, \label{projT} \ee projects vectors
onto the tangent 2-surfaces orthogonal to $e^{a}$ \textit{and}
$u^a$, which, following  \cite{extension}, we will refer to as `{\it
sheets}'. Hence it is obvious that $e^aN_{ab} = 0 =u^{a}N_{ab}$. In
(1+3) approach any second rank symmetric 4-tensor can be split into
a scalar along $u^a$, a 3-vector and a {\it projected symmetric
trace free} (PSTF) 3-tensor. In (1+1+2) slicing, we can take this
split further by splitting the 3-vector and PSTF 3-tensor with
respect to $e^a$. Any 3-vector, $\psi^{a}$, can be irreducibly split
into a component along $e^{a}$ and a sheet component $\Psi^{a}$,
orthogonal to $e^{a}$ i.e. \be \psi^{a} = \Psi e^{a} + \Psi^{a}\,,
\quad \Psi\equiv\psi^{a} e_{a}\,, \quad\Psi^{a} \equiv
N^{ab}\psi_{b}~. \label{equation1} \ee A similar decomposition can
be done for PSTF 3-tensor, $\psi_{ab}$, which can be split into
scalar (along $e^a$), 2-vector and 2-tensor part as follows: \be
\psi_{ab} = \psi_{\la ab\ra} = \Psi\bra{e_{a}e_{b} -
\frac{1}{2}N_{ab}}+ 2 \Psi_{(a}e_{b)} + \Psi_{ab}~,
\label{equation2} \ee where \bea
\Psi &\equiv & e^{a}e^{b}\psi_{ab} = -N^{ab}\psi_{ab}~,\nonumber \\
\Psi_{a} &\equiv & N_{a}{}^be^c\psi_{bc}~,\nonumber \\
\Psi_{ab} &\equiv & \psi_{\brac{{ab}}}
\equiv \bra{ N^{c}{}_{(a}N_{b)}{}^{d} - \frac{1}{2}N_{ab} N^{cd}} \psi_{cd}~,
\eea
and the curly brackets denote the PSTF part of a tensor with respect to $e^{a}$. \\
We also have
\be
h_{\left\{ab\right\}} = 0~,~ N_{\la ab\ra} = -e_{\la a}e_{b\ra} = N_{ab}- \frac{2}{3}h_{ab}~.
\ee
The sheet carries a natural 2-volume element, the alternating Levi-Civita
2-tensor:
\be
\veps_{ab}\equiv\veps_{abc}e^{c} = \eta_{dabc}e^{c}u^{d}~, \label{perm}
\ee
where $\veps_{abc}$ is the 3-space permutation symbol the volume element of the
3-space and $\eta_{abcd}$ is  the space-time permutator or the 4-volume.
With these definitions it follows that any 1+3 quantity  can be locally split in the
1+1+2 setting into only three types of objects: scalars, 2-vectors in the sheet,
and PSTF 2-tensors (also defined on the sheet).

Now apart from the `{\it time}' (dot) derivative of an object
(scalar, vector or tensor) which is the derivative along the
timelike congruence $u^a$, we now introduce two new derivatives,
which $ e^{a} $ defines, for any object $ \psi_{a...b}{}^{c...d}  $:
\bea \hat{\psi}_{a..b}{}^{c..d} &\equiv &
e^{f}D_{f}\psi_{a..b}{}^{c..d}~,
\label{hat}\\
\delta_f\psi_{a..b}{}^{c..d} &\equiv &
N_{a}{}^{f}...N_{b}{}^gN_{h}{}^{c}..
N_{i}{}^{d}N_f{}^jD_j\psi_{f..g}{}^{i..j}\;. \eea The hat-derivative
is the derivative along the $e^a$ vector-field in the surfaces
orthogonal to $ u^{a} $. The $\delta$ -derivative is the projected
derivative onto the orthogonal 2-sheet, with the projection on every
free index. We can now decompose the covariant derivative of $e^a$
in the direction orthogonal to $u^a$ into it's irreducible parts
giving \be {\rm D}_{a}e_{b} = e_{a}a_{b} + \frac{1}{2}\phi N_{ab} +
\xi\veps_{ab} + \zeta_{ab}~, \ee where \bea
a_{a} &\equiv & e^{c}{\rm D}_{c}e_{a} = \hat{e}_{a}~, \\
\phi &\equiv & \delta_ae^a~,\label{phi} \\  \xi &\equiv & \frac{1}{2}
\veps^{ab}\delta_{a}e_{b}~, \\
\zeta_{ab} &\equiv & \delta_{\lb a}e_{b \rb }~.
\eea
We see that along the spatial direction $e_a$, $\phi$ represents the
\textit{expansion of the sheet},  $\zeta_{ab}$ is the \textit{shear of $e^{a}$}
(i.e. the distortion of the sheet) and $a^{a}$ its \textit{acceleration}.
We can also interpret $\xi$ as the \textit{vorticity} associated with $e^{a}$
so that it is a representation of the ``twisting'' or rotation of the sheet.
The other derivative of $e^a$ is its change along $u^a$,
\be
\dot{e_a}=\A u_a +\alpha_a,
\ee
where we have $\A=e^a\dot{u_a}$ and $\alpha_a=N_{ac}\dot{e^c}$.
Also we can write the (1+3) kinematical variables and Weyl tensor as follows
\bea
\Theta&=&h_a^b\nab_bu^a\\
\dot{u}^{a} &=& \A e^{a}+ \A^{a}~,\label{1+1+2Acc} \\
\omega^{a} &=& \Omega e^{a} +\Omega^{a}~,
\eea
\bea
\sigma_{ab} &=& \Sigma\bra{ e_ae_b - \frac{1}{2}N_{ab}} +
2\Sigma_{(a}e_{b)} + \Sigma_{ab}~, \\
E_{ab} &=& \E\bra{ e_{a}e_{b} - \frac{1}{2}N_{ab}} +
2\E_{(a}e_{b)} + \E_{ab}~, \\
H_{ab} &=& \H\bra{ e_{a}e_{b} - \frac{1}{2}N_{ab}} + 2\H_{(a}e_{b)} +
\H_{ab}~.\label{MagH}
\eea
where $E_{ab}$ and $H_{ab}$ are the electric and magnetic part of the
Weyl tensor respectively. Therefore the key variables of the 1+1+2 formalism are
\bea
\left[ \Theta, \A, \Omega,\Sigma, \E, \H, \phi,\xi,
\A^{a},\Omega^{a}, \Sigma^{a}, \alpha^a, a^a,\E^{a}, \H^{a},\right. \nonumber\\
\left. \Sigma_{ab}, \E_{ab}, \H_{ab},\zeta_{ab} \right] \,.
\eea
These variables (scalars , 2-vectors and PSTF 2-tensors)
form an {\it irreducible set} and completely describe a vacuum spacetime.
In terms of these variables the full covariant derivatives of $e^{a}$ and $ u^{a}$ are

\bea
\nab_{a} e_{b} &=& - \A u_{a}u_{b} - u_{a}\alpha_{b} +
\bra{\Sigma + \frac13\Theta} e_{a} u_{b} \nonumber\\&&+ \bra{ \Sigma_{a} -
\veps_{ac}\Omega^{c}} u_{b}
+ e_{a}a_{b} + \frac12\phi N_{ab} \nonumber\\&&+ \xi\veps_{ab} + \zeta_{ab}~,
\label{eFullDerive}
\eea
\bea
\nab_{a}u_{b} &=& -u_{a}\bra{\A e_{b} + \A_{b}} +
e_{a}e_{b}\bra{ \frac13\Theta + \Sigma } \nonumber\\&&+ e_{a}\bra{ \Sigma_{b} +
\veps_{bc}\Omega^{c}}
+ \bra{\Sigma_{a}-\veps_{ac}\Omega^c}e_{b} \nonumber\\&&+ N_{ab}\bra{
\frac13\Theta - \frac12\Sigma} + \Omega \veps_{ab} + \Sigma_{ab}~.
\label{uFullDerive}
\eea

For the complete set of evolution equations, propagation equations,
mixed equations and constraints for the above irreducible set of variables
please see equations (48-81) of ~\cite{chris}.  Also we have the following commutation
relations for the different derivatives of any scalar $\psi$,
\be
\hat{\dot\psi}-\dot{\hat\psi}=-\A\dot\psi+(\frac13\Theta+\Sigma)\hat\psi+(\Sigma_a+
\veps_{ac} \Omega^c-\alpha_a)\delta^a\psi\;,
\label{com1}
\ee
\be
\delta_a\delta_b\psi-\delta_b\delta_a\psi=2\veps_{ab}(\Omega\dot\psi-\xi\hat\psi)
+2a_{[a} \delta_{b]}\psi
\label{com2}
\ee
From the above two relations it is clear that the 2-sheet is a genuine two surface
(rather than just a collection of tangent planes), in the sense that the commutator of the
time and hat derivative do not depend on any sheet component and also the sheet
derivatives commute, if and only if $\Sigma_a+
\veps_{ac} \Omega^c-\alpha_a=0$ and $\Omega=\xi=a^a=0$.

\section{Birkhoff Theorem for vacuum LRS-II spacetimes}
Locally Rotationally Symmetric (LRS) spacetimes posses a continuous
isotropy group at each point and hence a multi-transitive isometry
group acting on the spacetime manifold \cite{EllisLRS}. These
spacetimes exhibit locally (at each point) a unique preferred
spatial direction, covariantly defined. Since LRS spacetimes are
defined to be isotropic about a preferred direction, this allows for
the vanishing of \textit{all} orthogonal 1+1+2 vectors and tensors,
such that there are no preferred directions in the sheet. Then, all
the non-zero 1+1+2 variables are covariantly defined scalars. A
subclass of the LRS spacetimes, called LRS-II, contains all the LRS
spacetimes that are rotation free. As consequence in LRS-II
spacetimes the variables $\Omega$, $ \xi $ and $ \H $ are
identically zero and  the variables $\brac{\A, \Theta,\phi,
\Sigma,\E}$ fully characterize the kinematics of the vacuum
spacetime. The propagation and evolution equations of these
variables are: \bea \hat\phi&= -&\frac12\phi^2+\bra{\frac13\Theta
+\Sigma}\bra{\frac23\Theta-\Sigma}-\E\;, \label{LRS2start}\\
\hat\Sigma-\frac23\hat\Theta&= -&\frac32\phi\Sigma,\label{LRS2start1}\\
\hat\E&= -&\frac32\phi\E.\label{Ehat}
\eea
\bea
\dot\phi &= -&
\bra{\Sigma-\frac23\Theta}\bra{\A-\frac12\phi},\label{Q}\\
\dot\Sigma-\frac23\dot\Theta&= -&\A\phi +2\bra{\frac13\Theta
-\frac12\Sigma}^{2}-\E,\\
\dot\E&= &\bra{\frac32\Sigma-\Theta}\E. \label{LRS2middle} \eea \bea
\hat\A-\dot\Theta&= -&\bra{\A+\phi}\A+\frac13\Theta^2
+\frac32\Sigma^2. \label{LRS2end} \eea Since the vorticity vanishes,
the unit vector field $ u^{a} $ is hypersurface-orthogonal to the
spacelike 3-surfaces whose intrinsic curvature can be calculated
from the \textit{Gauss equation} for $ u^{a} $ that is generally
given as \cite{Gary}: \be ^{(3)}R_{abcd}=\bra{R_{abcd}}_{\bot} -
K_{ac}K_{bd} + K_{bc}K_{ad}\;, \ee where $ ^{(3)}R_{abcd} $ is the
\textit{3-curvature tensor}, $ \bot $ means projection with $ h_{ab}
$ on all indices and $ K_{ab} $ is the \textit{ extrinsic
curvature}. Also we note that in case of LRS-II spacetimes the
sheets at each point mesh together to form 2-surfaces. The Gauss
equation for $ e^{a} $ together with the 3-Ricci identities
determine the 3-Ricci curvature tensor of the spacelike 3-surfaces
orthogonal to $ u^{a} $ to be \be ^{3}R_{ab} =
-\bras{\hat{\phi}+\frac12 \phi^{2}}e_{a}e_{b} - \bras{\frac12
\hat{\phi} + \frac12\phi^{2} - K}N_{ab}\;. \ee This gives the
3-Ricci-scalar as \be ^{3}R = -2\bras{\frac12 \hat{\phi} +
\frac34\phi^{2} - K} \label{3nRicci} \ee where $ K $ is the
\textit{Gaussian curvature} of the sheet, $ ^{2}R_{ab}=KN_{ab} $ .
From this equation and (\ref{LRS2start}) an expression for $ K $ is
obtained in the form \cite{Gary} \be K = - \E + \frac14 \phi^{2} -
\bra{\frac13 \Theta - \frac12 \Sigma}^{2} \label{GaussCurv} \ee From
(\ref{LRS2start}-\ref{LRS2middle}), the evolution and propagation
equations of $K$ can be determined as \bea
\dot{K} = -\frac23 \bra{\frac23 \Theta - \Sigma}K,\label{evoGauss}\\
\hat{K} = -\phi K. \label{propGauss} \eea From equations
(\ref{Ehat}), (\ref{LRS2middle}),  (\ref{evoGauss}) and
(\ref{propGauss}) we get an interesting geometrical result for
vacuum LRS-II spacetime \be \E=CK^{3/2}. \label{EK} \ee That is, the
1+1+2 scalar of the electric part of the Weyl tensor is always
proportional to a power of the Gaussian curvature of the 2-sheet.
The proportionality constant $C$ sets up a scale in the problem. We
can immediately see that for Minkowski spacetime $C=0$.

To covariantly investigate the geometry of the vacuum LRS-II
spacetime, let us try to solve the {\it Killing equation} for a
Killing vector of the form $\xi_a=\Psi u_a+\Phi e_a$, where $\Psi$
and $\Phi$ are scalars. The Killing equation gives \be \nab_a(\Psi
u_b+\Phi e_b) + \nab_b(\Psi u_a+\Phi e_a) =0\;. \label{Killing} \ee
Using equations (\ref{eFullDerive}) and (\ref{uFullDerive}), and
multiplying the Killing equation by $u^au^b$, $u^ae^b$, $e^ae^b$ and
$N^{ab}$ we get the following differential equations and
constraints: \bea
\dot\Psi+\A\Phi&=&0\;, \label{psidot}\\
\hat\Psi -\dot\Phi-\Psi\A+\Phi(\Sigma+\frac13\Theta)&=& 0;,\label{psihat}\\
\hat\Phi+\Psi(\frac13\Theta+\Sigma)&=&0\;,\label{cons1}\\
\Psi(\frac23\Theta-\Sigma)+\Phi\phi&=&0\;.\label{cons2}
\eea
Now we know $\xi_a\xi^a=-\Psi^2+\Phi^2$. If $\xi^a$ is timelike (that is  $\xi_a\xi^a<0$),
then because of the arbitrariness in choosing the vector $u^a$, we can always make $\Phi=0$.
On the other hand, if $\xi^a$ is spacelike (that is  $\xi_a\xi^a>0$), we can make $\Psi=0$.

Let us first assume that $\xi^a$ is timelike and $\Phi=0$. In that
case we know that the solution of equations (\ref{psidot}) and
(\ref{psihat}) always exists while the constraints (\ref{cons1}) and
(\ref{cons2}) together imply that in general, (for a non trivial
$\Psi$), $\Theta=\Sigma=0$. Thus the expansion and shear of an unit
vector field along the timelike Killing vector vanishes. In this
case the spacetime is static. Now if $\xi^a$ is spacelike and
$\Psi=0$, solution of equations (\ref{psihat}) and (\ref{cons1})
always exists and the constraints (\ref{psidot}) and (\ref{cons2})
together imply that in general, (for a non trivial $\Phi$),
$\phi=\A=0$. From the LRS-II equations (\ref{LRS2start})-
(\ref{LRS2end}), we then immediately see that the spatial
derivatives of all quantity vanish and hence the spacetime is
spatially
homogeneous.\\

In other words, we can say: {\it There always exists a Killing vector in the local $[u,e]$ plane
for a vacuum LRS-II spacetime. If the Killing vector is timelike then the spacetime is locally
static, and if the Killing vector is spacelike the spacetime is locally spatially
homogeneous.}\\

In fact in the first case, when $\Theta=\Sigma=0$,  we have $\dot
K=0$. Furthermore if choose coordinates to make the Gaussian
curvature `$K$' of the spherical sheets proportional to the
inverse {square} of the radius co-ordinate `$r$', (such that this
coordinate becomes the {\it area radius} of the sheets), then this
geometrically relates the `{\it hat}' derivative with the radial
co-ordinate `$r$'. As we have already seen, $\hat{K} = -\phi K$,
where the hat derivative, defined in terms of the derivative with
respect to the co-ordinate `$r$', depends on the specific choice
of $e^a$ (orthogonal to $u^a$ and the sheet).  If we choose the
'radial' co-ordinate as the area radius of the spherical sheets,
then from (\ref{hat}) and (\ref{phi}) the hat derivative of any
scalar $M$ becomes \be \hat{M}=\frac{1}{2}r\phi\frac{d M}{d r}\,.
\label{hatr} \ee   for a static spacetime.  If the spacetime is
not static then there is also a dot derivative in the RHS of
equation (\ref{hatr}). Details are given in \cite{Gary}.

Now solving equations
(\ref{LRS2start})- (\ref{LRS2end}), we get the unique solution \bea
\phi=\frac{2}{r}\sqrt{1-\frac{2m}{r}}\;&,&\;\A=\frac{m}{r^2}
\left[1-\frac{2m}{r}\right]^{-\frac12}\\
\E=\frac{2m}{r^3}\;&,&\;K=\frac{1}{r^2}
\label{Schw1}
\eea
Here the constant $m$ is the constant of
integration. Solving for the metric components using the definition of these
geometrical quantities we get ~\cite{Gary}
\be
ds^2=-\left(1-\frac{2m}{r}\right)dt^2+\frac{dr^2}{(1-\frac{2m}{r})}+r^2d\Omega^2,
\ee
which is the metric of a static Schwarzschild exterior.
In the second case, when $\phi=\A=0$, we can choose $u^a=\sqrt{\frac{2m}{t}-1}\delta^a_t$
where $m$ is a constant
and then solving (\ref{LRS2start})- (\ref{LRS2end}), we get the unique solution
\bea
\Theta=\frac{3m-2t}{t\sqrt{t(2m-t)}} \;&,&\;\Sigma=-\frac23\frac{3m-t}{t\sqrt{t(2m-t)}},\\
\E=-\frac{2m}{t^3}\;&,&\;K=\frac{1}{t^2}
\label{Schw2}
\eea
Again solving for the metric components we get
\be
ds^2=-\frac{dt^2}{(\frac{2m}{t}-1)}+\left(\frac{2m}{t}-1\right)dr^2+t^2d\Omega^2,
\ee
which is a part of the Schwarzschild solution inside the Schwarzschild radius.\\

Thus we have proved the (local) {\bf Birkhoff Theorem}: {\it Any
$C^2$ solution of Einstein's equations in empty space, which is of
the class LRS-II in an open set ${\mathcal{S}}$, is locally
equivalent to part of maximally extended Schwarzschild solution is
${\mathcal{S}}$.} Also it is interesting to note that the modulus of
the proportionality constant in equation (\ref{EK}), which sets a
scale in the problem, is exactly equal to the Schwarzschild radius.

\section{Almost Birkhoff Theorem}

As we have already seen, for a spherically symmetric spacetime, the
1+1+2 scalar of the electric part of the Weyl tensor is always
proportional to the $(3/2)^{th}$ power of the Gaussian curvature of
the 2-sheet, with the proportionality constant defining a scale in
the problem. Let us now perturb this geometry and define the notion
of an {\it almost spherically symmetric} spacetime
in the following way:\\

{\it Any $C^2$ spacetime that admits a local 1+1+2 splitting at
every point such that the magnitude of all 2-vectors on the sheet,
the sheet gradient of scalars (defined by $\sqrt{\psi_a\psi^a}$),
the magnitude of all PSTF 2-tensors on the sheet, and the sheet
derivative of 2-vectors (defined by $\sqrt{\psi_{ab}\psi^{ab}}$)
at any given point are either zero or much smaller than the scale
defined by the modulus of the proportionality constant in equation
(\ref{EK}), is called an \emph{almost spherically symmetric} spacetime. }\\

We would like to emphasize here that though Minkowski spacetime
belongs to the set of LRS-II, in the above definition of the
perturbed spacetime we exclude the Minkowski background, as in that
case the scale is identically zero. As we have seen from equations
(\ref{com1}) and (\ref{com2}), the sheet will be a genuine two
surface if and only if the commutator of the time and hat derivative
do not depend on any sheet component and also the sheet derivatives
commute. In the perturbed scenario we will require the sheet to be
an almost genuine 2-surface in the sense that the commutator of the
time and hat derivative almost do not depend on any sheet component
and the sheet derivatives almost commute; this will follow from our
definition of almost spherically symmetric. In that case the scalars
$\Omega$ and $\xi$ would be of the same order of smallness as the
other vectors and PSTF 2-tensors on the sheet. Also using the
constraint equation
 \be
\delta_a\Omega^a +\veps_{ab}\delta^a\Sigma^b=(2\A-\phi)\Omega-3\xi\Sigma
+\veps_{ab}\zeta^{ac}\Sigma^b_c+\H\;,
\ee
we see the scalar $\H$ also is of the same order of smallness. Hence the set
of 1+1+2 variables
\bea
\left[ \Omega, \H, \xi,
\A^{a},\Omega^{a}, \Sigma^{a}, \alpha^a, a^a, \right. \nonumber\\
\left. \E^{a}, \H^{a}, \Sigma_{ab}, \E_{ab}, \H_{ab},\zeta_{ab}
\right] \,. \eea are all of ${\mathcal O}(\epsilon)$ with respect to
the invariant scale. Using  equations (48-81) of ~\cite{chris}, we
can get the propagation and evolution equations of these small
quantities. We now list all the equations up to the first order
(that is up to order `$\epsilon$'). The time evolution equations of
$\xi$ and $\zeta_{\lb ab\rb}$ are as follows: \be \dot\xi =
\bra{\f12\Sigma-\f13\theta}\xi +\bra{\udota-\f12\phi}\Omega
+\f12\lc_{ab}\delta^a\dotn^b+\f12\H\;, \label{dotxinl} \ee \bea
\dot\zeta_{\lb ab\rb}&=&\bra{\f12\Sigma-\f13\theta}\zeta_{ab}
+\bra{\udota-\f12\phi}\Sigma_{ab} \nonumber\\&&+\delta_{\lb
a}\alpha_{b\rb} -\lc_{c\lb a}\H_{b\rb}^{~~c}\;. \label{dotzetanl}
\eea The Vorticity evolution equations: \bea
\dot\Omega&=&\f12\lc_{ab}\delta^a\udota^b+\udota\xi+\Omega\bra{\Sigma-\f23\theta}\;,
\eea \bea
\dot\Omega_{a}+\f12\lc_{ab}\hat\udota^b&=&-\bra{\f23\theta+\f12\Sigma}\Omega_a
\nonumber\\&&+\f12\lc_{ab}\bras{-\udota
    a^b+\delta^b\udota-\f12\phi\udota^b}\;.
\eea
Shear evolution equations:
\bea
\dot\Sigma_{\lb ab\rb}&=&\delta_{\lb a}\udota_{b\rb}
+\udota\zeta_{ab}-\bra{\f23\theta+\f12\Sigma}\Sigma_{ab}-\E_{ab}\;,
\eea
\bea
\dot\Sigma_{a}-\f12\hat\udota_{a}&=&\f12\delta_a\udota+\bra{\udota-\f14\phi}\udota_a
    -\bra{\f23\theta+\f12\Sigma}\Sigma_a\nonumber\\&&
    +\f12\udota a_a-\f32\Sigma\alpha_a-\E_a
\eea
Evolution equation for $\hat e_a$:
\bea
\hat\alpha_{a}-\dot a_{a}&=&
    -\bra{\f12\phi+\udota}\alpha_a
    +\bra{\f13\theta+\Sigma}\bra{\udota_a-a_a}\nonumber\\&&
    +\bra{\f12\phi-\udota}\bra{\Sigma_a+\lc_{ab}\Omega^b}
   -\lc_{ab}\H^b.\label{hatalphanl}
\eea
Electric Weyl evolution:
\bea
\dot\E_{a}+\:12\lc_{ab}\hat\H^b&=&
    \f34\lc_{ab}\delta^b\H+\f12\lc_{bc}\delta^b\H^c_{~a}
    -\f34\E\Sigma_a\nonumber\\&&
    +\f34\E\lc_{ab}\Omega^b
    -\f32\E\alpha_a
    +\bra{\f34\Sigma-\theta}\E_a \nonumber\\&&
    -\bra{\f14\phi+\udota}\lc_{ab}\H^b\;,
\eea
\bea
\dot\E_{\lb ab\rb}-\lc_{c\lb a}\hat\H_{b\rb}^{~~c}&=&
    -\lc_{c\lb a}\delta^c\H_{b\rb}
    -\f32\E\Sigma_{ab}
    -\bra{\theta+\f32\Sigma}\E_{ab} \nonumber\\&&
    +\bra{\f12\phi+2\udota}\lc_{c\lb a}\H_{b\rb}^{~~c}\;.
\eea
Magnetic Weyl evolution:

\bea
\dot\H&=&-\lc_{ab}\delta^a\E^b-3\xi\E
    +\bra{\theta+\f32\Sigma}\H\;,
\eea

\bea
\dot\H_{a}-\f12\lc_{ab}\hat\E^b&=&
     \bra{\f34\Sigma-\theta}\H_a
    -\f32\E\lc_{ab}\udota^b\nonumber\\&&
    +\f34\E\lc_{ab}a^b-\f12\lc_{bc}\delta^b\E^c_{~a}\nonumber\\&&
    +\bra{\f14\phi+\udota}\lc_{ab}\E^b
   -\f34\lc_{ab}\delta^b\E\;,
\eea

\bea
\dot\H_{\lb ab\rb}+\lc_{c\lb a}\hat\E_{b\rb}^{~~c}&=&
    +\f32\E\lc_{c\lb a}\zeta_{b\rb}^{~~c}
    -\bra{\f12\phi+2\udota}\lc_{c\lb a}\E_{b\rb}^{~~c}\nonumber\\&&
    -\bra{\theta+\f32\Sigma}\H_{ab}+\lc_{c\lb a}\delta^c\E_{b\rb}\;.
\eea

In the above equation all the zeroth order quantities are background quantities.
If the background is static with $\Theta=\Sigma=0$ and the time derivative
all the background quantities are zero, we can easily see that the time derivatives of
the first order quantities at a given point is of the same order of smallness as
themselves. Hence the first order quantities still remains  ``small'' as the time evolves.

Similarly we can write the spatial propagation equation of all the
first order quantities up to  ${\mathcal O}(\epsilon)$. The
propagation equations of $\xi$ and $\zeta_{\lb ab\rb}$ are: \bea
\hat\xi&=&-\phi\xi+\bra{\f13\theta+\Sigma}\Omega+\f12\lc_{ab}\delta^aa^b\;,
\eea \bea \hat\zeta_{\lb ab\rb}&=&-\phi\zeta_{ab} +\delta_{\lb
a}\hatn_{b\rb } +\bra{\f13\theta +\Sigma}\Sigma_{ab}-{\cal
E}_{ab}\;.\label{hatzetanl} \eea Shear divergence: \bea
\hat\Sigma_{a}-\lc_{ab}\hat\Omega^b&=&\f12\delta_a\Sigma
    +\f23\delta_a\theta - \lc_{ab}\delta^b\Omega-\f32\phi\Sigma_a
     -\f32\Sigma a_a\nonumber\\&&
    +\bra{\f12\phi+2\udota}\lc_{ab}\Omega^b
    -\delta^b\Sigma_{ab}\;,
\eea
\bea
\hat\Sigma_{\lb ab\rb}&=&\delta_{\lb a}\Sigma_{b\rb} -\lc_{c\lb a}\delta^c\Omega_{b\rb}
    -\f12\phi\Sigma_{ab}\nonumber\\&&
    +\f32\Sigma\zeta_{ab}-\lc_{c\lb a}\H_{b\rb}^{~~c}\;.
\eea
Vorticity divergence equation:
\bea
\hat\Omega&=&-\delta_a\Omega^a+\bra{\udota-\phi}\Omega\;.\label{hatOmSnl}
\eea
Electric Weyl Divergence:
\bea
\hat\E_{a}&=& \f12\delta_a\E -\delta^b\E_{ab}
      -\f32\E a_a-\f32\phi\E_a\;.
\eea
Magnetic Weyl divergence:
\bea
\hat\H&=& -\delta_a\H^a
    -\f32\phi\H-3\E\Omega\;,
\eea
\bea
\hat\H_{a}&=& \:12\delta_a\H-\delta^b\H_{ab}
    -\f32\E\lc_{ab}\Sigma^b
    +\f32\E\Omega_a\nonumber\\&&
    +\f32\Sigma\lc_{ab}\E^b
    -\f32\phi\H_a\;.
\eea Hence we see that if the background is spatially homogeneous
with $\phi=\A=0$ and the `hat' derivative all the background
quantities are zero, we can easily see that the `hat' derivatives of
the first order quantities at a given point are of the same order of
smallness as themselves. Hence the first order quantities still
remain  ``small'' along the spatial direction. In both the cases of
a static background and a spatially homogeneous background the
resultant set of equations are the perturbed LRS-II equations, (that
is equations (\ref{LRS2start})- (\ref{LRS2end}) with ${\mathcal
O}(\epsilon)$ terms added to each).

Again trying to solve the {\it Killing equation} (\ref{Killing}) for a
Killing vector of the form $\xi_a=\Psi u_a+\Phi e_a$,
using equation (\ref{uFullDerive}),(\ref{eFullDerive})
and multiplying the Killing equation by $u^au^b$,
$u^ae^b$, $e^ae^b$, $N^{ab}$, $N^a_cu^b$, $N^a_ce^b$ and $N^a_cN^b_d$,
we get the following differential equations and constraints:
\bea
\dot\Psi+\A\Phi&=&0\;, \label{psidot1}\\
\hat\Psi -\dot\Phi-\Psi\A+\Phi(\Sigma+\frac13\Theta)&=& 0\;,\label{psihat1}\\
\hat\Phi+\Psi(\frac13\Theta+\Sigma)&=&0\;,\label{cons11}\\
\Psi(\frac23\Theta-\Sigma)+\Phi\phi&=&0\;,\label{cons21} \\
-\delta_c\Psi+\Psi\A_c+\Phi\bra{\veps_{cd}\Omega^d+\alpha_c+\Sigma_c}&=&0\;,\label{delpsi}\\
\delta_c\Phi+\Phi a_c +2\Psi\Sigma_c&=&0\;,\label{delphi}\\
\Psi\Sigma_{cd}+\Phi\zeta_{cd}&=&0\;.\label{cons3} \eea Now we see
that for both timelike ($\Phi=0$)  or spacelike ($\Psi=0$)
vectors, all the above equations are not completely solved in
general unless the first order quantities appearing in the
equations above are exactly equal to zero. This special case
corresponds to static but distorted black holes in the presence of
matter outside black hole, for example when an accretion disk
occurs ~\cite{Frolov}. If the distribution of the matter outside
black hole is axisymmetric, then the vacuum metric outside the
matter is described by Weyl Solution. However as we proved that
these first order quantities generically remain ${\mathcal
O}(\epsilon)$ both in space and time, we can see that a timelike
vector with ($\Theta=\Sigma=0$) or a spacelike vector with
($\phi=\A=0$)
almost solves the Killing equations. Therefore we can say:\\

{\it For an almost spherically symmetric vacuum spacetime there
always exists a vector in the local $[u,e]$ plane which almost
solves the Killing equations. If this vector is timelike then the
spacetime is locally almost static, and if the Killing vector is
spacelike the spacetime is locally almost spatially homogeneous.}\\

Also as we have seen that in this case the resultant set of
equations are the perturbed LRS-II equations, (that is equations
(\ref{LRS2start})- (\ref{LRS2end}) with ${\mathcal O}(\epsilon)$
terms added to each), and the perturbations locally remain small
both in space and time, a part of the maximally extended
almost-Schwarzschild solution will then solve the field equations locally.\\

Thus we have proved the (local) {\bf Almost Birkhoff Theorem}:
{\it Any $C^2$ solution of Einstein's equations in empty space,
which is almost spherically symmetric in an open set
${\mathcal{S}}$, is locally almost equivalent to part of a
maximally extended Schwarzschild solution in ${\mathcal{S}}$.}\\

Note that we do not consider perturbations across the horizon: our
result holds for any open set ${\mathcal{S}}$ that does not
intersect the horizon in the background spacetime. The result almost
certainly holds true across the horizon also, but that case needs
separate consideration.\\

The above result can be immediately generalized in the presence of a
cosmological constant. In that case an `almost' spherically symmteric
solution in an open set ${\mathcal{S}}$, is
locally almost equivalent to part of a maximally extended
Schwarzschild deSitter/anti-deSitter solution in  ${\mathcal{S}}$.
Also the result holds for an almost
spherically symmetric electric charge distribution with no spin or
magnetic dipole. In this case we have to use the energy momentum
tensor of the electromagnetic field in vacuum with the magnetic part
being equal to zero. Proceeding exactly in a similar manner one can
then show that the solution of the perturbed field equations will be
almost equivalent to a part of maximally extended
Reissner-Nordstr$\rm{\ddot{o}}$m spacetime.

\section{Discussion}
The rigidity of spherical vacuum solutions of the EFE, as enshrined
in Birkhoff's theorem, is maintained in the perturbed case: almost
spherical symmetry implies almost static. This is an important
reason for the stability of the solar system, and of black hole
spacetimes.

Though there are many discussions in the literatures on the
stability of Schwarzschild solution in General Relativity, for
example \cite{stability}; most of them deal with a specific sector
of the maximally extended Schwarzschild manifold, namely the static
exterior part. In this paper we have established in a compact and
completely different way, an aspect of the stability of the static
sector of the complete manifold: \emph{as long as the solution
remains almost spherically symmetric, it remains almost static},
with a similar result for the spatially homogeneous sector.
Furthermore our result, being local, does not depend on specific
boundary conditions used for solving the perturbation equations.
Hence it does not depend on the global topology of the spacetime,
and brings out covariantly the rigidity and uniqueness of the
almost-spherical vacuum solutions of Einstein's field equations.

This rigidity has interesting implications for the issue of how a
universe made up of locally spherically symmetric objects imbedded
in vacuum regions is able to expand, given that Birkhoff's theorem
tells us the local spacetime domains have to be static. A two-mass
exact solution illustrating this situation is given in
\cite{twomass}; the present paper suggests the results given there
are stable.


\end{document}